\RequirePackage[2020-02-02]{latexrelease}
\documentclass[aps,11pt,preprintnumbers,nofootinbib,floatfix]{revtex4}

\usepackage{subfigure}
\usepackage{multirow}
\usepackage{amsmath}

\usepackage{amssymb}
\usepackage{graphicx}

\begin{document}

\title{Topological equivalence and phase transition rate in holographic thermodynamics of regularized Maxwell theory}

\author{Tran N. Hung}
\email{hung.tranngoc@phenikaa-uni.edu.vn}  
\affiliation{Phenikaa Institute for Advanced Study and Faculty of Fundamental Sciences, Phenikaa University, Yen Nghia, Ha Dong, Hanoi 12116, Vietnam}
\author{Cao H. Nam}
\email{nam.caohoang@phenikaa-uni.edu.vn}  
\affiliation{Phenikaa Institute for Advanced Study and Faculty of Fundamental Sciences, Phenikaa University, Yen Nghia, Ha Dong, Hanoi 12116, Vietnam}
\date{\today}

\begin{abstract}
Utilizing the holographic dictionary from the proposal that treats Newton's constant as a thermodynamic variable, we establish a thermodynamic topological equivalence between the AdS black holes in the bulk and the thermal states in the dual CFT. The findings further reveal that the thermodynamic topological characteristics of the RegMax AdS black holes are strongly influenced by the characteristic parameter of the regularized Maxwell theory. Additionally, we investigate the phase transition between low and high entropy thermal states within a canonical ensemble in the dual CFT. Our observations indicate that the phase transition behavior of the thermal states mirrors that of the black holes. By modeling the phase transition process as a stochastic process, we are able to calculate the rates of phase transition between the thermal states. This result enhances our understanding of the dominant processes involved in the phase transition of the thermal states in the dual CFT.
\end{abstract}

\maketitle

\section{Introduction}

In recent decades, theoretical studies on black holes have shifted from solely gravitational considerations to exploring their microscopic nature, with the goal of uncovering insights into the theory of quantum gravity. Black hole thermodynamics, initiated by the groundbreaking works of Jacob Bekenstein and Stephen Hawking in the 1970s, is one of the most active fields in this area. Employing a semi-classical approach, Hawking derived the notable result that black holes emit thermal radiation, referred to as Hawking radiation, due to vacuum fluctuations near the event horizon \cite{Hawk1975}. He also proposed the area theorem, which states that the area of a black hole's event horizon can never decrease \cite{Hawk1972}. Inspired by this theorem, Bekenstein conjectured that black hole entropy is proportional to the area of the event horizon \cite{Beken1972}. Furthermore, Bardeen, Carter, and Hawking established the four laws of black hole thermodynamics \cite{Bacaha1973}. Accordingly, the first law of black hole thermodynamics is formulated as follows
\begin{eqnarray}
    \delta M=T\delta S+\Omega \delta J+\Phi \delta Q
\end{eqnarray}
for a black hole of mass $M$, charge $Q$, and angular momentum $J$. However, this equation does not contain a pressure-volume term $V\delta P$ which typically presents in standard thermodynamics. In recent years, this problem prompted the idea of the negative cosmological constant acting as the positive pressure of the black hole via the following relationship \cite{Dolan2011}
\begin{eqnarray}
    P=-\frac{\Lambda}{8\pi G_N}=\frac{(D-1)(D-2)}{16\pi G_N l^2}, 
\end{eqnarray}
where $G_N$ stands for the Newton's constant and $l$ is the $D$ dimension AdS curvature radius. The thermodynamic volume is identified as the conjugate variable of pressure. By incorporating these new quantities, the extended form of the first law of thermodynamics and the corresponding Smarr relation for a charged AdS black hole can be expressed as follows \cite{Kumate2017}
\begin{eqnarray}
    \delta M&=&T\delta S+V\delta P+\Omega \delta J+\Phi \delta Q,\\
    M&=&\frac{D-2}{D-3}(TS+\Omega J)+\phi Q-\frac{2}{D-3}PV.
\end{eqnarray}
In this new framework, the mass $M$ has no longer meaning of internal energy but reinterpreted as a chemical enthalpy \cite{Karatra2009}. In the extended phase space, the thermodynamics of these AdS black holes exhibit a wide range of phenomena, including Van der Waals-like phase transitions \cite{Kuma2012,Caca2013,Nam2018}, reentrant phase transitions \cite{Alkuma2013,Kurihe2021,Cuxuzhu2023}, black hole heat engine \cite{John2014,Bhaye2017,Zhali2019,Nam2021}, and Joule-Thomson expansions \cite{Okay2017,Molila2018,Cihu2019,Nam2020}. 

Despite these interesting results, the extended phase space thermodynamics lacks a clear holographic interpretation. In the anti-de Sitter/conformal field theory (AdS/CFT) correspondence, the thermodynamics of the gravitational theory in asymptotically spacetime is equivalent to that of the dual CFT \cite{Witt1998,Karatra2009}. Therefore, we expect that the thermodynamic variables of AdS black holes should be dual to standard thermodynamic variables in the CFT. However, the variable $\Lambda$ is not clearly interpretable in the context of holographic duality. A variation of $\Lambda$ corresponds to changing both the central charge $C$ (or the number of colors $N$) and the CFT volume $\mathcal{V}$. Consequently, the first law of AdS black holes cannot be straightforwardly related to the CFT first law \cite{Karob2015,Cokuma2021}.

To reestablish the extended phase space thermodynamics for holographic consistency, Visser proposed considering Newton's constant $G_N$ as a thermodynamic variable \cite{Visser2022}. Incorporating variations of $\Lambda$ and $G_N$, the extended first law of charged AdS black holes, along with the generalized Smarr relation, are expressed as follows
\begin{eqnarray}
    dM&=&\frac{\kappa}{8\pi G_N}dA+\Omega dJ+\Phi dQ+\frac{\Theta}{8\pi G_N}d\Lambda-(M-\Omega J-\Phi Q)\frac{dG_N}{G_N},\label{eqn:first1}\\
    M&=&\frac{D-2}{D-3}\left(\frac{\kappa A}{8\pi G_N}+\Omega J\right)+\Phi Q-\frac{1}{D-3}\frac{\Theta\Lambda}{4\pi G_N},
\end{eqnarray}
where $\Theta=-V$ is the conjugate quantity to the cosmological constant, which can be defined as $\int_{\Sigma_{BH}}|\xi|dV-\int_{\Sigma_{AdS}}|\xi|dV$, with $|\xi|$ being the norm of the time translation Killing vector $\xi$ \cite{Kara2009,Javi2019}. In this definition the domain $\Sigma_{BH}$ extends from the horizon to infinity, and $\Sigma_{AdS}$ in the pure AdS integral extends across the entire spacetime.

We observe that the $\Lambda$ and $G_N$ variations in (\ref{eqn:first1}) cannot be combined into a single term $d(\Lambda/G_N)$. This indicates that the standard interpretation of the extended first law in terms of the pressure $P=-\Lambda/8\pi G_N$ is inconsistent when Newton's constant varies. This inconsistency can be resolved by considering the central charge $C$ as a thermodynamic variable in the dual CFT. According to the AdS/CFT correspondence, the central charge of the dual CFT is related to both the AdS radius and Newton's constant, $C\sim l^{D-2}/G_N$. Therefore, variations of $C$ in the CFT could lead to variations of both $\Lambda$ and $G_N$ in the bulk.

Using Smarr relation and inserting $d\Lambda/\Lambda=-2dl/l$, we can rewrite the extended first law as
\begin{eqnarray}
    dM&=&\frac{\kappa}{2\pi}d\left(\frac{A}{4G_N}\right)+\Omega dJ+\frac{\Phi}{l}d(Ql)-\frac{M}{D-2}\frac{dl^{D-2}}{l^{D-2}}\nonumber\\
    &+& \left(M-\frac{\kappa A}{8\pi G_N}-\Omega J-\frac{\Phi}{l}Ql\right)\frac{d(l^{D-2}/G_N)}{l^{D-2}/G_N}.
\end{eqnarray}
Inserting the standard holographic dictionary for CFTs living on the boundary, the extended CFT first law is expressed as follows \cite{Visser2022}
\begin{eqnarray}
    dE=\mathcal{T}d\mathcal{S}-pd\mathcal{V}+\phi d\mathcal{Q}+\mu dC+\Omega d\mathcal{J},
\end{eqnarray}
where $E,p$, and $\mathcal{V}$ are the CFT energy, pressure, and volume, respectively. The advantage of this new rewriting is that all the terms in the first law have a thermodynamic interpretation in the dual CFT.

Models of non-linear electrodynamics (NLEs) have long been investigated as classical attempts to regularize the field of a point charge in Maxwell's theory \cite{Boin1933,Soro2022}. The important idea was to modify the electromagnetic Lagrangian so that the point charge could exhibit a finite field strength and finite self-energy. Among these, perhaps the most renowned NLE theory is the Born-Infeld theory \cite{Boin1933,Boin1934}. This theory features finite self-energy for point charges, possesses electromagnetic duality, and approximates Maxwell's electrodynamics in the weak field limit. It was later discovered that Born-Infeld-type Lagrangians also emerge in the low-energy regime of string theory \cite{Fratse1985} and in D-brane physics \cite{Leigh1989}. Recently, Einstein gravity coupled with NLEs has garnered significant scholarly attention due to its potential in addressing the singularity problem associated with black holes \cite{Bardeen1968,Beagar2000,Fanwa2016}. Interestingly, in the extended phase space, the nonlinear electrodynamic (NLE) AdS black holes also possesses thermodynamic behaviors such as Van der Waals-like phase transitions \cite{Guku2012,Kuli2018,Mabre2020} and Joule-Thomson expansions \cite{Bame2022,Krug20221,Krug20222,Krug2023}.

In this paper, we explore the intriguing characteristics of the holographic thermodynamics of AdS black hole solutions within the regularized Maxwell (RegMax) theory, a NLE theory first introduced in \cite{Taham2021}. This theory, which admits the Maxwell limit, offers the simplest regularization of the point charge field and its self-energy. Aside from Maxwell's theory, it stands as a unique model of non-linear electrodynamics that provides radiative solutions in the Robinson-Trautman class. Notably, slowly rotating black holes or charged accelerated black holes within a given NLE framework, first constructed in the ModMax theory \cite{Baci2022}, are also naturally found in the RegMax theory \cite{Taham2021,Haku2023}. The spherical black hole solutions and their thermodynamic properties within RegMax theory were thoroughly investigated in \cite{Haku2023}. Their findings demonstrate that the characteristic parameter of RegMax theory significantly influences the thermodynamic behavior of the black holes.

This paper is organized as follows. In the next section, we summarized interesting features of RegMax AdS black holes. Sect. III provides an overview of the established holographic thermodynamics and defines the thermodynamic quantities of the dual thermal states of RegMax AdS black holes. In Sect. III.A, we demonstrate a general result that AdS black holes exhibit topological thermodynamic equivalence between the bulk and the dual CFT by employing the generalized free energy landscape. Detailed calculations of the winding numbers in the topological thermodynamic space of RegMax AdS black holes are provided. Sect. III.B investigates the phase transition features in canonical ensembles of the thermal states in the dual CFT. We observe that the thermodynamic behavior of the thermal states is strongly dependent on the characteristic parameter in RegMax theory. We identify Van der Waals-like behavior in the phase transition and determine the phase transition rates using the thermal potential framework \cite{Zmxu2021} and Kramer's escape rate from stochastic processes \cite{Zwan2001}. Finally, we make conclusions in Sect. IV.

\section{RegMax AdS black holes}
In the NLE theories, the non-linear electrodynamics is minimally coupled to Einstein's gravity described by the following action \cite{Haku2023}
\begin{eqnarray}
    I=\frac{1}{16\pi G_N}\int d^4 x\sqrt{-g}(R+4\mathcal{L}-2\Lambda),
\end{eqnarray}
where $\Lambda$ is the cosmological constant. $\mathcal{L}$ is the non-linear electromagnetic Lagrangian which is defined as a function of the two electromagnetic invariants
\begin{eqnarray}
    \mathcal{S}=\frac{1}{2}F_{\mu\nu}F^{\mu\nu},\quad \mathcal{P}=\frac{1}{2}F_{\mu\nu}(*F^{\mu\nu}),
\end{eqnarray}
where the electromagnetic field strength $F_{\mu\nu}=\partial_{\mu}A_{\nu}-\partial_{\nu}A_{\mu}$. RegMax theory belongs to a restricted class of NLEs, whose Lagrangian is only $\mathcal{L}(\mathcal{S})$ as follows
\begin{eqnarray}
    \mathcal{L}&=&-2\alpha^4\left(1-3\log (1-s)+\frac{s^3+3s^2-4s-2}{2(1-s)}\right),\\
    s&\equiv& \left(-\frac{\mathcal{S}}{\alpha^4}\right)^{\frac{1}{4}}\in (0,1).
\end{eqnarray}
The theory is characterized by a parameter $\alpha>0$ that has dimension (length)$^{-1/2}$. In the weak field approximation, $\alpha\rightarrow\infty$, the theory reduces to the Maxwell theory due to $\mathcal{L}\rightarrow \mathcal{L}_M=-\frac{1}{2}\mathcal{S}$.

The RegMax AdS black hole is characterized by the static spherically symmetric metric reads
\begin{eqnarray}
    ds^2=-f(r)dt^2+\frac{dr^2}{f(r)}+r^2 d\Omega^2,
\end{eqnarray}
where the metric function as
\begin{eqnarray}
    f(r)&=&1-2\alpha^2 |Q|+\frac{4\alpha |Q|\sqrt{|Q|}-6 m}{3r}+4r\alpha^3\sqrt{|Q|}\\ \nonumber
    & &-4\alpha^4 r^2 \log\left(1+\frac{\sqrt{|Q|}}{r\alpha}\right)+\frac{r^2}{l^2}.
\end{eqnarray}
The solution exhibits a singularity at $r=0$, characterized by divergent Ricci and Kretschmann scalars as follows
\begin{eqnarray}
    R&=&-\frac{4|Q|\alpha^2}{r^2}+O(1/r),\\
    \mathcal{K}&=&R_{\alpha\beta\gamma\sigma}R^{\alpha\beta\gamma\sigma}\nonumber\\
    &=&\frac{16(2\alpha|Q|^{3/2}-3m)^2}{3r^6}+O(1/r^5).
\end{eqnarray}
Nonetheless, the severity of the Kretschmann scalar's divergence can be mitigated by selecting specific values for the mass parameter $m$ and the charge $Q$. 

The values of parameters determine whether the solution is a black hole with one or two horizons, or a naked singularity. We can expand the metric function around the origin as follows
\begin{eqnarray}
    f=\frac{2(M_m - m)}{r}+1-2\alpha^2  |Q|+4\alpha^3 r\sqrt{|Q|}+O(r^2),
\end{eqnarray}
where
\begin{eqnarray}
    M_m=\frac{2\alpha |Q| \sqrt{|Q|}}{3}.
\end{eqnarray}
For $m>M_m$, the solution exhibits a single horizon, characteristic of a Schwarzschild-like (S-type) black hole. Conversely, for $m<M_m$, the solution manifests as a Reissner–Nordström-like (RN-type) black hole. In the marginal case where $m=M_m$, the metric function attains a finite value at the origin, $f_0(r=0)=1-2|Q|\alpha^2$. If this function is positive, the RN-type solution is a naked singularity.

The thermodynamic mass $M$ can be defined as the mass parameter $m$
\begin{eqnarray}
    M=m &=&\frac{2}{3}|Q|\sqrt{|Q|}\alpha+\frac{1-2 |Q|\alpha^2}{2}r_h+2\sqrt{|Q|}\alpha^3 r_{h}^{2}\\ \nonumber
    & &+\frac{r_{h}^{3}}{2 l^2}-2\alpha^4r_{h}^{3}\log\left(1+\frac{\sqrt{|Q|}}{\alpha r_h}\right).
\end{eqnarray}
The temperature and entropy are calculated by the standard formulae and read
\begin{eqnarray}
    T&=&\frac{f'(r_h)}{4\pi}\\ \nonumber
    &=&\frac{2}{\pi}\sqrt{|Q|}\alpha^3+\frac{1-2|Q|\alpha^2}{4\pi r_h}+\frac{3r_h}{4\pi l^2}\\ \nonumber
    & &+\frac{\sqrt{ |Q|}\alpha^4 r_h}{\pi(\sqrt{ |Q|}+\alpha r_h)}+\frac{3}{\pi}\alpha^4 r_h \log\left(1+\frac{\sqrt{ |Q|}}{\alpha r_h}\right),\\
    S&=&\frac{\pi r_{h}^{2}}{G_N}.
\end{eqnarray}
Because the solution is asymptotically an AdS black hole, we can define the pressure and conjugate volume as \cite{Kumate2017}
\begin{eqnarray}
    P&=&-\frac{\Lambda}{8\pi G_N},\\
    V&=&\left(\frac{\partial M}{\partial P}\right)=\frac{4}{3}\pi r_{h}^{3}.
\end{eqnarray}
From these quantities, the first law in the thermodynamic extended phase space is written as
\begin{eqnarray}
    dM=TdS+\Phi dQ+VdP+\mu_{\alpha} d\alpha.
\end{eqnarray}
This is accompanied by the Smarr relation
\begin{eqnarray}
    M=2TS+\Phi Q-2VP-\frac{1}{2}\mu_{\alpha} \alpha,
\end{eqnarray}
where the parameter $\alpha$ is considered as an $\alpha-$polarization potential.

\section{Topological and holographic thermodynamics in RegMax theory}
In the context of the AdS/CFT correspondence, the dual field theory resides on the conformal boundary of the asymptotically AdS bulk spacetime. According to the prescription provided in \cite{Gukpo1998}, the CFT metric is defined by the AdS metric evaluated on the boundary, adjusted by a Weyl factor. Consequently, the line element of the CFT can be expressed as \cite{Cokuma2022}
\begin{eqnarray}
    ds^2=\frac{R^2}{l^2}(-dt^2+R^2 d\Omega_{2}^{2}),
\end{eqnarray}
where $d\Omega_{2}^{2}$ is the differential element of a 2-sphere and $R$ is the curvature radius of the boundary.
The spatial volume of the boundary sphere is defined by the boundary curvature radius as follows
\begin{eqnarray}
    \mathcal{V}=\Omega_2 R^2,
\end{eqnarray}
where $\Omega_2$ is the volume of the unit 2-sphere. For this choice of CFT metric, the holographic dictionary is written as follows
\begin{eqnarray}\label{eqn:dictionary}
    E&=&M\frac{l}{R},\quad \mathcal{T}=\frac{\kappa}{2\pi}\frac{l}{R},\quad \mathcal{S}=S=\frac{A}{4G_N},\quad C=\frac{l^2}{4G_N}\\ \nonumber
    \phi&=&\frac{\Phi}{l}\frac{l}{R},\quad \mathcal{Q}=Q l,\quad \tilde{\mu}_{\alpha}=\frac{\mu_{\alpha}}{l}\frac{l}{R},\quad \tilde{\alpha}=\alpha l.
\end{eqnarray}
Using this holographic dictionary, we can infer the extended first law and Euler relation for the boundary CFT as follows
\begin{eqnarray}
    dE&=&\mathcal{T}d\mathcal{S}+\phi d\mathcal{Q}+\tilde{\mu}_{\alpha}d\tilde{\alpha}-pd\mathcal{V}+\mu dC,\\
    E&=&\mathcal{T}\mathcal{S}+\phi \mathcal{Q}-\frac{1}{2}\tilde{\mu}_{\alpha}\tilde{\alpha}+\mu C.
\end{eqnarray}
The pressure and the chemical potential are defined as
\begin{eqnarray}
    p&=&\frac{E}{2\mathcal{V}},\\
    \mu&=&\frac{1}{C}\left(E-\mathcal{T}\mathcal{S}-\phi \mathcal{Q}+\frac{1}{2}\tilde{\mu}_{\alpha}\tilde{\alpha}\right).
\end{eqnarray}
Finally, we explicitly determine the thermodynamic variables of the dual CFT. For convenience, we introduce the dimensionless parameters
\begin{eqnarray}\label{eqn:parachange}
    x\equiv \frac{r_h}{l},\quad y\equiv \alpha\sqrt{l}.
\end{eqnarray}
Using the holographic dictionary, we can define the energy, the temperature and the entropy of the CFT as follows
\begin{eqnarray}
    E&=&\frac{1}{R}\left[2Cx+2Cx^3+\frac{|\mathcal{Q}|}{3}\sqrt{\frac{|\mathcal{Q}|}{C}}y-|\mathcal{Q}|xy^2+4\sqrt{|\mathcal{Q}|C}x^2y^3 \right.\\ \nonumber
    &&\left.-8Cx^3y^4\log\left(1+\sqrt{\frac{|\mathcal{Q}|}{C}}\frac{1}{2xy}\right) \right],\\
    \mathcal{T}&=&\frac{1}{4\pi R}\left[3x+\frac{1}{x}-\frac{|\mathcal{Q}|}{2C}\frac{y^2}{x}+4\sqrt{\frac{|\mathcal{Q}|}{C}}y^3+2\sqrt{\frac{|\mathcal{Q}|}{C}}y^3\left(1+\sqrt{\frac{|\mathcal{Q}|}{C}}\frac{1}{2xy}\right)^{-1}\right.\label{eqn:temp}\\ \nonumber
    & &\left.-12xy^4 \log\left(1+\sqrt{\frac{|\mathcal{Q}|}{C}}\frac{1}{2xy}\right) \right],\\
    \mathcal{S}&=&4\pi C x^2.\label{eqn:entropy}
\end{eqnarray}

\subsection{Thermodynamic topology equivalence between bulk and boundary}
In topological thermodynamic spaces, the black hole and the corresponding boundary CFT states are considered as defects. Utilizing Duan’s topological current theory, we can determine the local and global topological winding numbers of these defects \cite{Weilima2022}. In this section, we establish that the topological numbers are identical in the bulk and the boundary. In other words, the thermodynamic states of the black holes are topologically equivalent to the thermodynamic states of the CFTs.

Prior to specific calculations, we wish to make a few remarks on the topological current method in the thermodynamic topology of black holes. We define a vector field $\phi$, using the generalized free energy $F=M-\frac{S}{\tau}$, as follows
\begin{eqnarray}
    \phi=(\phi^{1},\phi^{2})\equiv\left(\frac{\partial F}{\partial r_h},-\cot \Theta \csc \Theta\right),
\end{eqnarray}
where the parameter $0\leq \Theta \leq \pi$ for convenience. The defects, referred to as zero points, correspond to the conditions $\Theta=\frac{\pi}{2}$ and $\frac{\partial F}{\partial r_h}=0$, which is equivalent to $\tau=T^{-1}$, where $T$ represents the black hole temperature. The topological current as
\begin{eqnarray}
    j^{\mu}=\frac{1}{2\pi}\epsilon^{\mu\nu\rho}\epsilon_{ab}\partial_{\nu}n^{a}\partial_{\rho}n^{b},\quad \mu, \nu, \rho=0,1,2,
\end{eqnarray}
where $\partial_{\nu}=\frac{\partial}{\partial x^{\nu}}$ with $x^{\nu}=(\tau,r_h,\Theta)$ and the unit vector $n^{a}=\frac{\phi^{a}}{||\phi^{a}||}$ $(a=1,2)$. Using the Jacobi tensor $\epsilon^{ab}J^{\mu}\left(\frac{\phi}{x}\right)=\epsilon^{\mu\nu\rho}\partial_{\nu}\phi^{a}\partial_{\rho}\phi^{b}$, we can reexpress the current as
\begin{eqnarray}
    j^{\mu}=\delta^{2}(\phi)J^{\mu}\left(\frac{\phi}{x}\right).
\end{eqnarray}
The winding numbers of the $i$-th zero point $z_i$ can be defined as $\omega_i=\beta_i \eta_i$, where the positive Hopf index $\beta_i$ counts the number of loops that $\phi^a$ makes in the vector space when $x^\mu$ goes around the zero point $z_i$, and the Brouwer degree $\eta_i=$ sign$(J^{0}|_{z_i})=\pm 1$. From the definition of the Jacobi tensor, we find that
\begin{eqnarray}
    J^{0}|_{z_i}=\partial_{1}\phi^{1}\partial_{2}\phi^{2}|_{z_i}.
\end{eqnarray}
On the other hand, $\Theta$ has the value of $\frac{\pi}{2}$ at the zero points, thus $\partial_2 \phi^2 |_{z_i}=1$. Therefore, the winding number $\omega_i$ can be simply calculated as follows
\begin{eqnarray}
    \omega_i=\text{sign}(J^{0}|_{z_i})=\text{sign}(\partial_1 \phi^1|_{z_i})=\text{sign}\left(\frac{\partial^2 F}{\partial r_{h}^{2}}|_{z_i}\right).
\end{eqnarray}
Based on the preceding analysis, we can infer that the zero points align with the extrema of the generalized free energy. Specifically, local maxima, associated with unstable black holes, have a winding number of -1, while local minima, corresponding to metastable or stable black holes, exhibit a winding number of 1. The global topological winding number $W$ is derived by summing all the winding numbers $\omega_i$
\begin{eqnarray}
    W=\sum_{i=1}^{N} \omega_i.
\end{eqnarray}
Hence, under the assumption of smoothness for the generalized free energy, the global winding number $W$ only takes three values, -1, 0, and 1, in accordance with the conjecture proposed in \cite{Weilima2022}. 

In holographic thermodynamics of the canonical ensemble, we define two generalized free energies for the bulk and the dual CFT, respectively, as follows
\begin{eqnarray}
    F&=&M-\frac{S}{\tau},\\
    \mathcal{F}&=&E-\frac{\mathcal{S}}{\tilde{\tau}},
\end{eqnarray}
where the parameters $\tau$ and $\tilde{\tau}$ can be considered as the inverse temperatures of the ensembles. The holographic dictionary (\ref{eqn:dictionary}) gives us the equality $\mathcal{F}=F\frac{l}{R}$. For the grand canonical ensemble, the generalized free energies $W=M-\frac{S}{\tau}-\Phi Q$ and $\mathcal{W}=E-\frac{\mathcal{S}}{\tilde{\tau}}-\phi\mathcal{Q}$ also have the same relation, $\mathcal{W}=W\frac{l}{R}$. Because the two generalized free energies differ by only one factor, we can conclude, based on previous remarks, that there is thermodynamic topological equivalence between the bulk and the CFT.

\begin{figure}[ht]
  \begin{subfigure}
  \centering
    \includegraphics[width=0.4\linewidth]{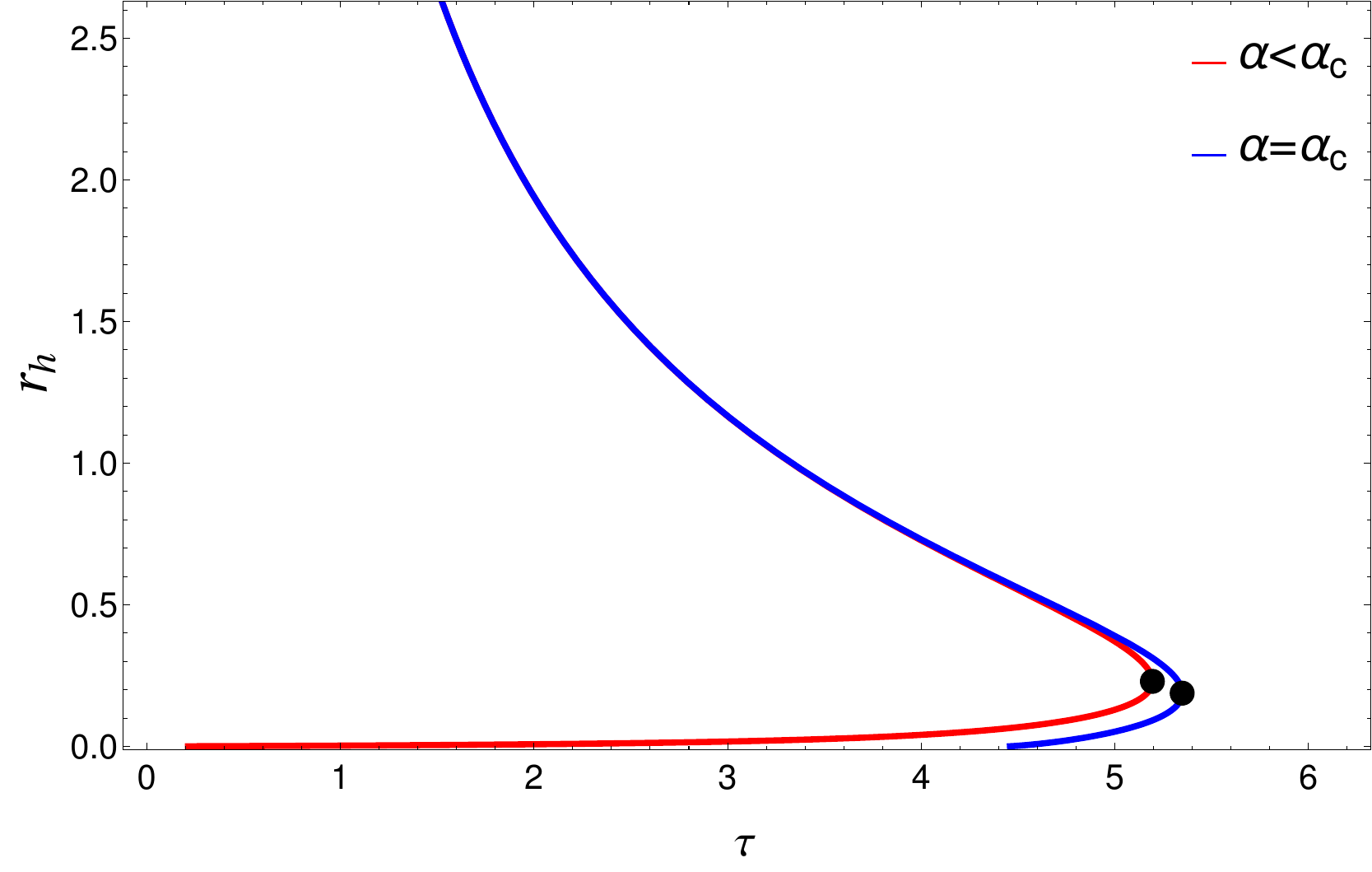}
  \end{subfigure}%
  \begin{subfigure}
  \centering
    \includegraphics[width=0.4\linewidth]{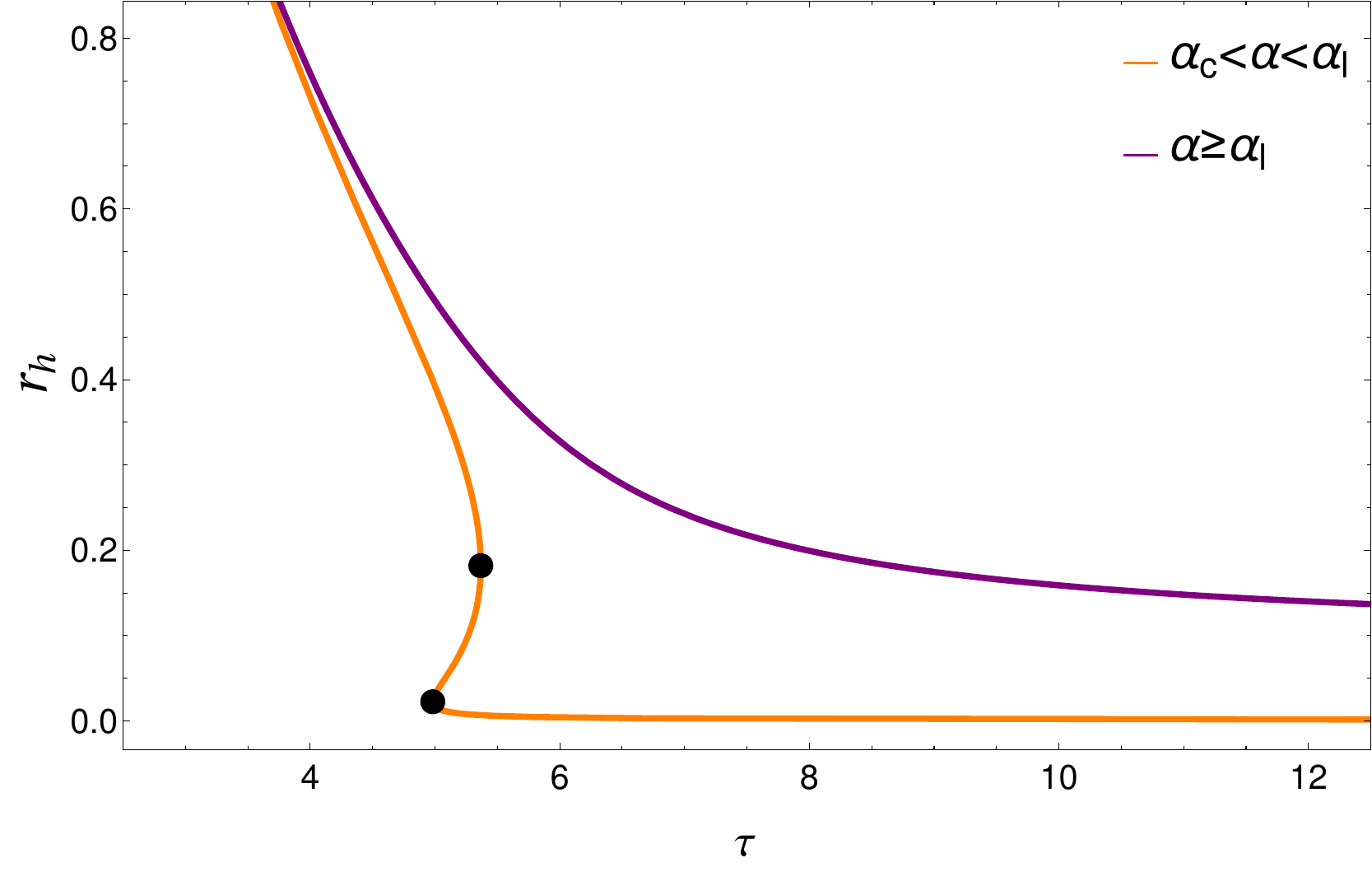}
  \end{subfigure}
  \caption{The zero-point curves are shown in the $\tau-r_h$ plane. The black dots represent the reversal positions of the curves. They divide the curves into different branches: small, intermediate, and large black holes.}
  \label{fig:zeropoint}
\end{figure}

The extended phase space thermodynamics of the RegMax AdS black holes, encompassing both the canonical and grand canonical ensembles, was investigated in \cite{Haku2023}. Notably, the phase transition properties exhibit a strong dependence on the dimensionful parameter $\alpha$, particularly when it crosses the critical value. In this section, we aim to elucidate how the thermodynamic topology of these solutions varies with changes in this parameter. To achieve this, we will determine the topological winding numbers of the RegMax AdS black hole in detail. The generalized free energy of the black hole in the canonical ensemble is given by
\begin{eqnarray}
    F&\equiv& M-\frac{S}{\tau}\\ \nonumber
    &=&\frac{2}{3}|Q|\sqrt{|Q|}\alpha+\frac{1-2|Q|\alpha^2}{2}r_h+2\sqrt{|Q|}\alpha^3 r_{h}^{2}\\ \nonumber
    & &+\frac{r_{h}^{3}}{2 l^2}-2\alpha^4r_{h}^{3}\log\left(1+\frac{\sqrt{|Q|}}{\alpha r_h}\right)-\frac{\pi r_{h}^{2}}{G_N\tau}.
\end{eqnarray}
For simplicity we fix other parameters as $l=1$, $G_N=1$, and $Q=1$. From the condition $\frac{\partial F}{\partial r_h}=0$, we can obtain the parameter $\tau$ of the zero points $z_i$ as follows
\begin{eqnarray}\label{eqn:tau}
    \tau=-\frac{4\pi r_h(1+r_h\alpha)}{1-2\alpha^2+(\alpha+6\alpha^3)r_h+3(1+4\alpha^4)r_{h}^{2}+3\alpha r_{h}^{3}-12\alpha^4(1+\alpha r_h)r_{h}^{2}\log\left[1+\frac{1}{\alpha r_h}\right]}.
\end{eqnarray}
We find some interesting properties of $\tau$ following this formula. The denominator has one positive solution $r_h>0$ if $\alpha> \frac{1}{\sqrt{2}}$, and it is always positive if $\alpha< \frac{1}{\sqrt{2}}$. From equation (\ref{eqn:tau}) we obtain $\tau \rightarrow 0$ when $r_h\rightarrow 0$ in both cases. Additionally, if $\alpha=\alpha_c=\frac{1}{\sqrt{2}}$, $\tau\rightarrow \pi\sqrt{2}$ as $r_h\rightarrow 0$. Through numerical calculations, we also determine that $\tau$ has two local extrema if $\frac{1}{\sqrt{2}}<\alpha<\alpha_l\simeq 0.713$ and has no extrema if $\alpha>\alpha_l$. The zero-point curves are depicted in the $\tau-r_h$ plane, as shown in Fig. \ref{fig:zeropoint}.

\begin{figure}[ht]
  \begin{subfigure}
  \centering
    \includegraphics[width=0.3\linewidth]{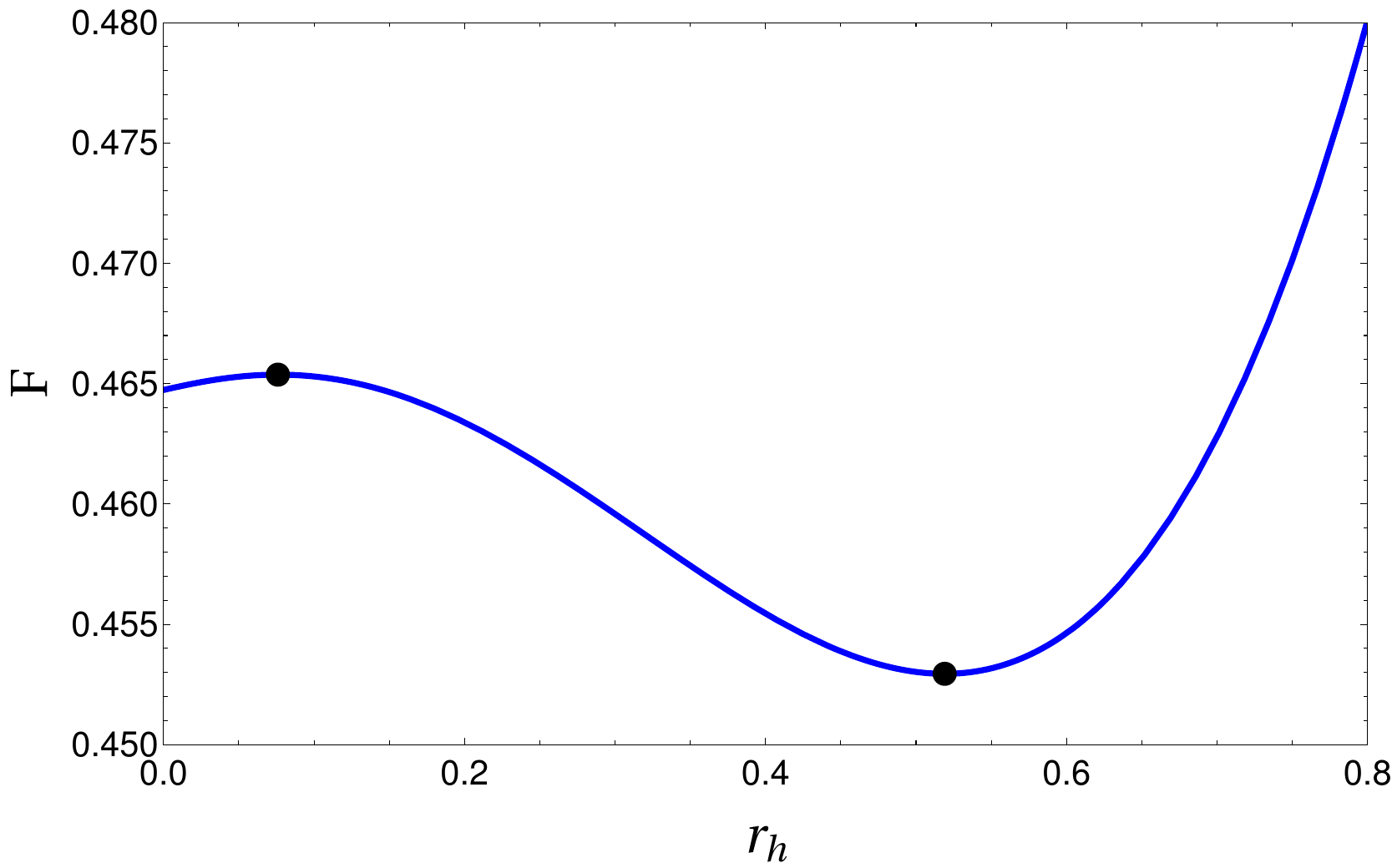}
  \end{subfigure}%
  \begin{subfigure}
  \centering
    \includegraphics[width=0.3\linewidth]{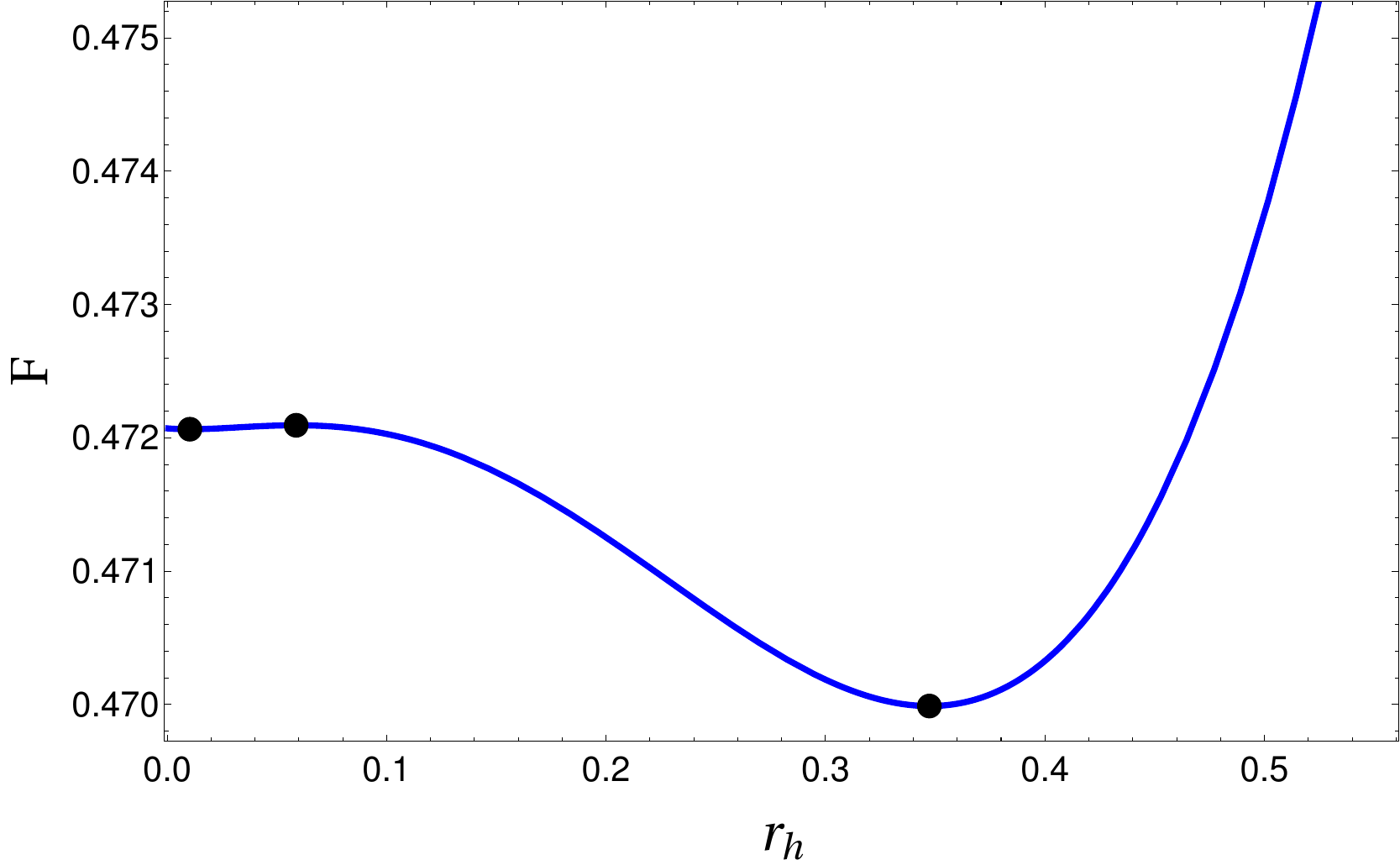}
  \end{subfigure}
  \begin{subfigure}
  \centering
    \includegraphics[width=0.3\linewidth]{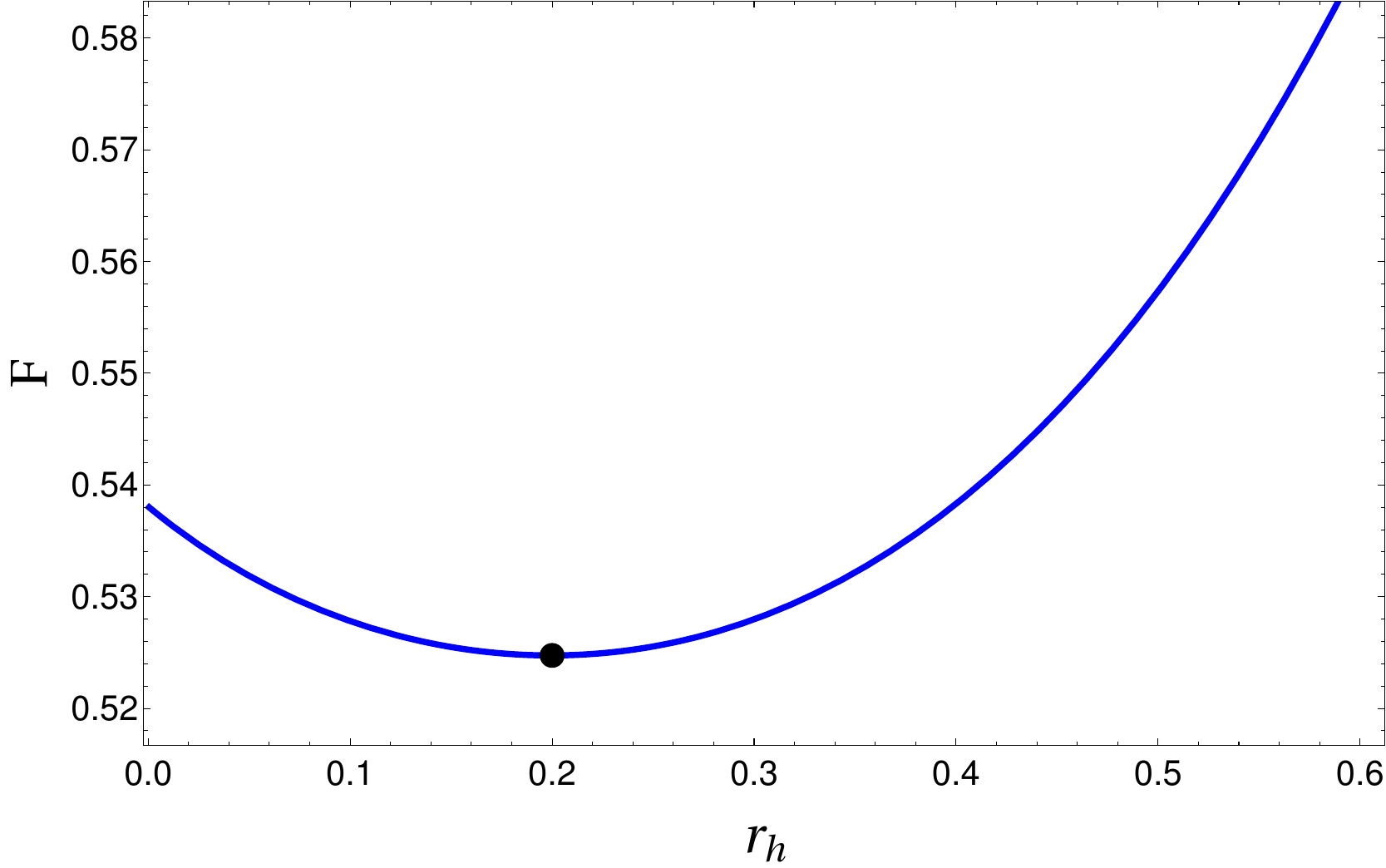}
  \end{subfigure}
  \caption{The generalized free energy $F$ versus the horizon radius $r_h$. The left, middle, and right panel for $\alpha\leq \alpha_c$, $\alpha_c<\alpha<\alpha_l$, and $\alpha\geq \alpha_l$, respectively.}
  \label{fig:gfe}
\end{figure}

For $\alpha\leq\alpha_c$, the collection of zero points in the $\tau-r_h$ plane forms curves as illustrated in the left panel of Fig. \ref{fig:zeropoint}. There exists a maximum value $\tau_m$ (black dots) corresponding to the minimum temperature of the black hole, $T_m$. For $\tau<\tau_m$ (or $T>T_m$), the generalized free energy $F$ behaves as shown in the left panel of Fig. \ref{fig:gfe}. Two black dots represent two extrema, corresponding to two zero points at a constant $\tau$ in the $\tau-r_h$ plane. The stable black hole, corresponding to the minimum, has a positive winding number 1, while the unstable black hole, corresponding to the maximum, has a negative winding number -1. Thus, the global winding number of the topological thermodynamic space is zero. 

For values of $\alpha$ greater than the critical value $\alpha_c$, the zero-point curves are depicted in the right panel of Fig. \ref{fig:zeropoint}. Notably, no phase transition occurs when $\alpha$ is greater than or equal to $\alpha_l$. However, in the range $\alpha_c<\alpha<\alpha_l$, the curves exhibit two inflection points corresponding to distinct values of $\tau$, $\tau_1<\tau_2$. These inflection points mark changes in the direction of variation of $\tau$. In the case $\tau_1<\tau<\tau_2$, the generalized free energy $F$ is illustrated in the middle panel of Fig. \ref{fig:gfe}. Specifically, the behavior of $F$ reveals two local minima and one local maximum. The first local minimum, observed at small horizon radii, characterizes a metastable black hole. Conversely, the second local minimum, associated with large horizon radii, corresponds to a stable black hole. The local maximum, situated at intermediate horizon radii, describes an unstable black hole that rapidly transitions to more stable states. This behavior bears similarities to the Van der Waals fluid. Finally, we can compute the local winding numbers for the metastable, unstable, and stable black holes, yielding values of 1, -1, and 1, respectively. Thus, the global winding number is equal to 1.

\subsection{Holographic thermodynamics in the dual CFT}
Employing the holographic dictionary, we study the extended thermodynamics of the thermal states in the dual CFT. We examine the phase transition in a canonical ensemble with fixed values of $(\mathcal{Q}, \mathcal{V}, C)$, considering its dependence on the parameter $y$, which is related to the characteristic parameter $\alpha$ of the regular Maxwell theory via Equation (\ref{eqn:parachange}). Utilizing Kramer's escape rate from stochastic processes, we calculate the phase transition rates of the thermal states within the thermal potential framework proposed in \cite{Zmxu2021}.

\subsubsection{Phase transition in canonical ensemble}

In the canonical ensemble with fixed $(\mathcal{Q},\mathcal{V},C)$, we have the free energy as
\begin{eqnarray}
    G&\equiv& E-\mathcal{T}\mathcal{S}\\ \nonumber
    &=&\frac{1}{R}\left[Cx-Cx^3+\frac{\mathcal{Q}}{3}\sqrt{\frac{\mathcal{Q}}{C}}y-\frac{\mathcal{Q}}{2}xy^2-2\sqrt{\mathcal{Q}C}x^2 y^3 \left(1+\sqrt{\frac{\mathcal{Q}}{C}}\frac{1}{2xy}\right)^{-1}\right.\\ \nonumber
    &&\left. +4Cx^3 y^4\log\left(1+\sqrt{\frac{\mathcal{Q}}{C}}\frac{1}{2xy}\right)\right]
\end{eqnarray}
In this section, we study the dependence of the free energy $G$ on the parameter $y$, which is directly related to $\alpha$ via Equation (\ref{eqn:parachange}), while keeping $R=1$, $C=1$, and $\mathcal{Q}=1$. For this scenario, the temperature $\mathcal{T}$ and the free energy $G$ are re-expressed as follows
\begin{eqnarray}
    \mathcal{T}&=&\frac{1}{4\pi}\left[3x+\frac{2-y^2}{2x}+4y^3+2y^3\left(1+\frac{1}{2xy}\right)^{-1}-12xy^4\log\left(1+\frac{1}{2xy}\right)\right],\\
    G&=&x-x^3+\frac{1}{3}y-\frac{1}{2}xy^2-2x^2 y^3\left(1+\frac{1}{2xy}\right)^{-1}+4x^3 y^4\log\left(1+\frac{1}{2xy}\right).
\end{eqnarray}

\begin{figure}[ht]
  \centering
    \includegraphics[width=0.5\linewidth]{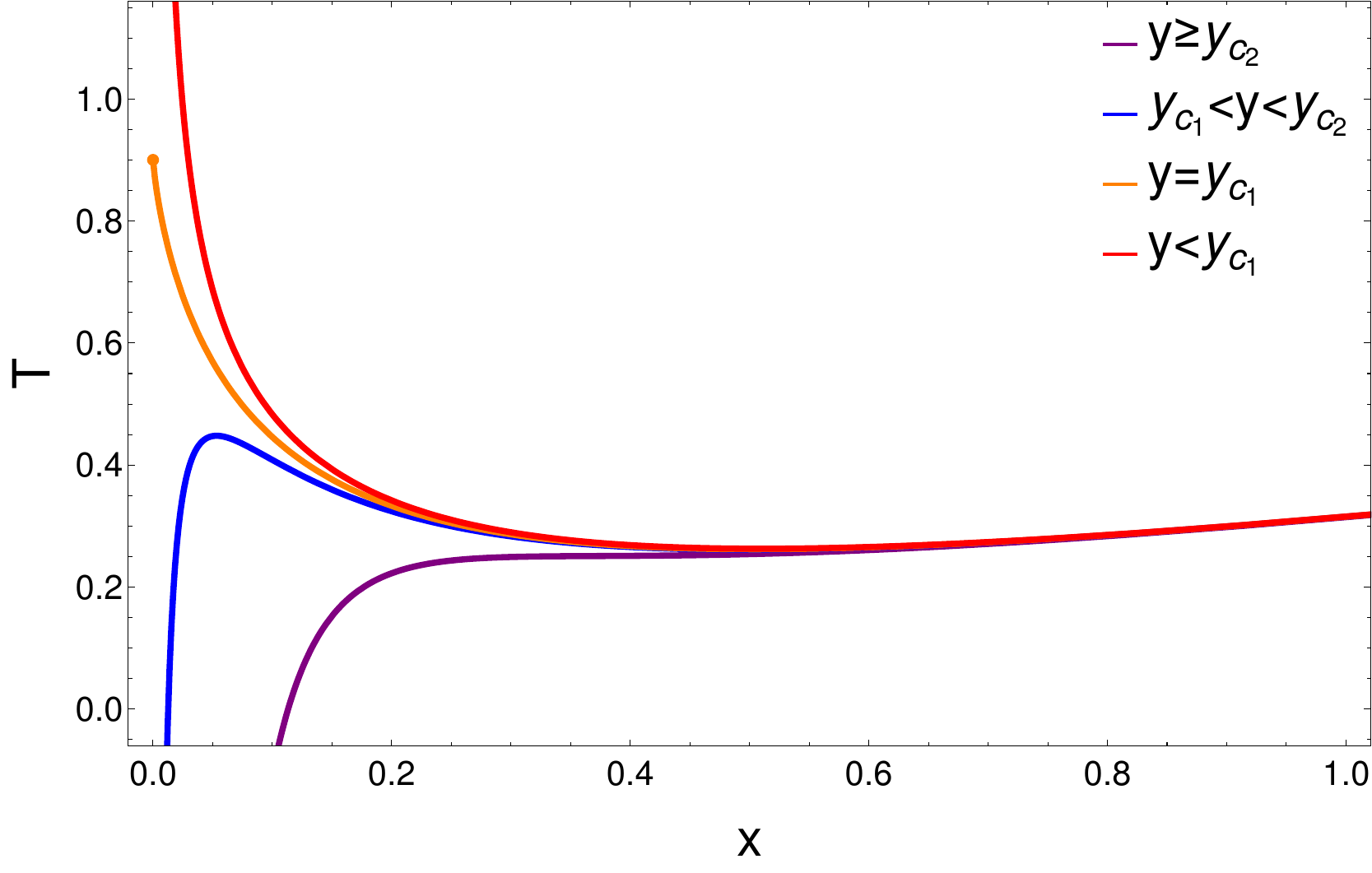}
  \caption{The $\mathcal{T}-x$ curves of the canonical ensemble for different values of $y$.}
  \label{fig:temp}
\end{figure}

The behavior of the temperature of the thermal state is depicted by the $\mathcal{T}-x$ curves for different values of $y$ in Fig. (\ref{fig:temp}). We observe that the curves exhibit the same behavior for large $x$. However, they become distinct for small $x$. At the critical value $y=y_{c_1}=\sqrt{2}$, the temperature curve approaches a definite value of $\frac{2\sqrt{2}}{\pi}$ as $x\rightarrow 0$. For $y<y_{c_1}$, the temperature becomes positively indefinite as $x\rightarrow 0$. For $y>y_{c_1}$, the temperature curves approach zero at specific values of $x$. In this case, we identify another critical value $y_{c_2}\approx 2.795$ at which the following system of equations has a solution
\begin{eqnarray}
    \frac{\partial \mathcal{T}}{\partial x}=0, \quad \frac{\partial^2 \mathcal{T}}{\partial x^2}=0.
\end{eqnarray}
For $y_{c_1}<y<y_{c_2}$, the temperature curves exhibit two local extrema, which may be associated with a Van der Waals-like phase transition. A clearer understanding of this phase transition can be achieved by examining the behavior of the free energy.

\begin{figure}[ht]
  \centering
    \includegraphics[width=0.5\linewidth]{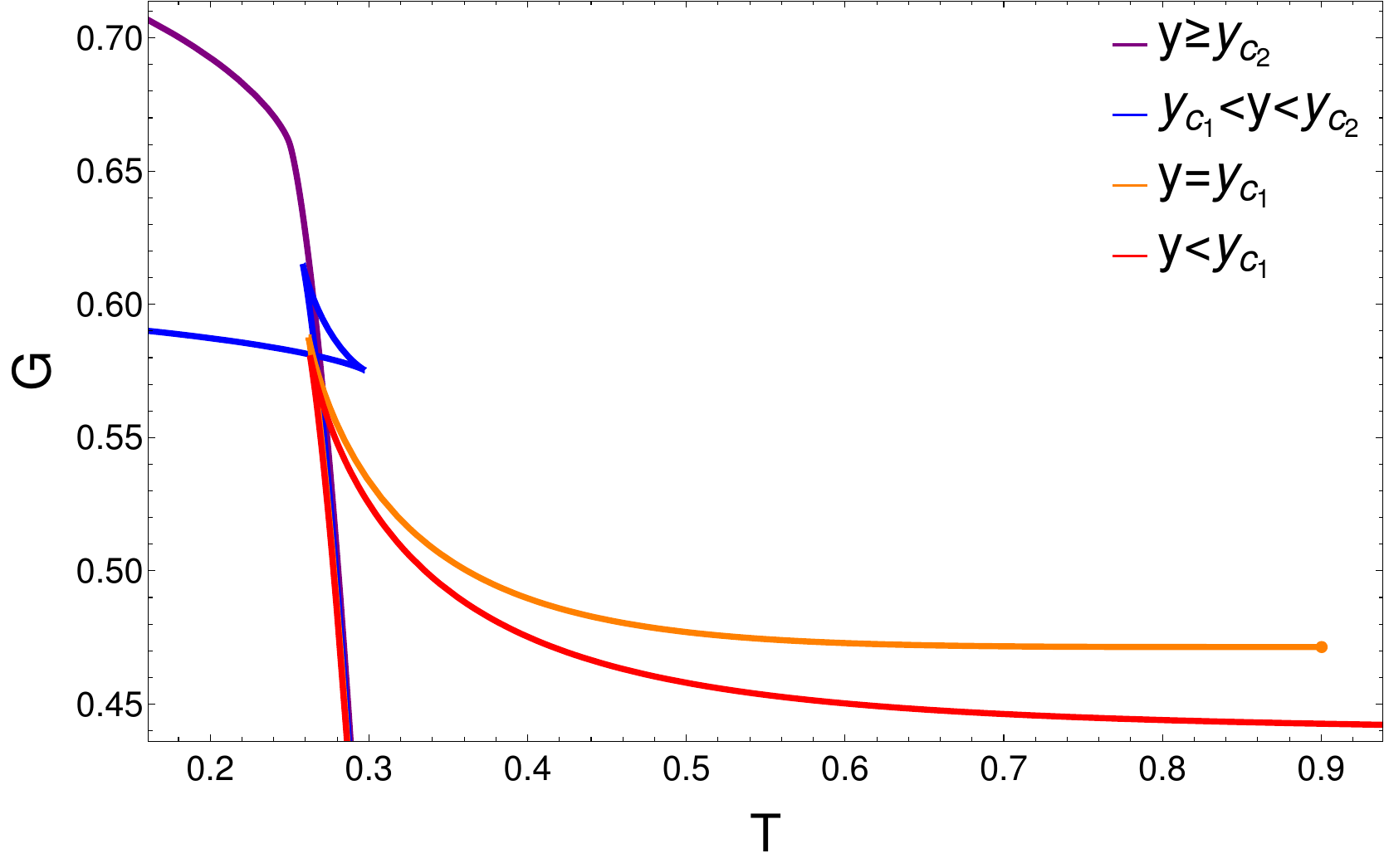}
  \caption{The free energy $G$ versus the horizon radius $T$.}
  \label{fig:gibbs}
\end{figure}

The free energy versus temperature curves are illustrated in Fig. (\ref{fig:gibbs}). For $y_{c_1}<y<y_{c_2}$, the free energy exhibits a swallowtail shape. Along the curve, the value of $x$ increases from the leftmost point. According to Equation (\ref{eqn:entropy}), small black holes correspond to low entropy thermal states in the dual CFT, while large black holes correspond to high entropy thermal states. Therefore, the self-intersection point of the curve signifies a first-order phase transition between low and high entropy thermal states. The phase transition rate for this case will be calculated in the next section. When $y>y_{c_2}$, the curve of the free energy becomes monotonic, indicating a single-phase state. For $y\leq y_{c_1}$, the curve exhibits a cusp at a minimum temperature. At this cusp, a second-order phase transition occurs  between the low entropy states (in the upper branch) and high entropy states (in the lower branch). We observe that the behavior of the free energy of the thermal states is similar to that of the RegMax AdS black holes studied in \cite{Haku2023}.

\subsubsection{Phase transition rates}
We consider a canonical ensemble comprising a large number of states, including the thermal states in the dual CFT. The phase transition can be understood as the rearrangement of thermal states in the ensemble caused by thermal fluctuations. Using Kramer escape rate from stochastic processes \cite{Zwan2001,Xuwu2023} and the thermal potential framework described in \cite{Zmxu2021}, we investigate the phase transition rates between the low and high entropy thermal states in the case of swallowtail free energy.  The thermal potential of the canonical ensemble is defined as follows \cite{Zmxu2021}
\begin{eqnarray}
    U=\int \left(\mathcal{T}-t\right)d\mathcal{\mathcal{S}},\label{eqn:therpot}
\end{eqnarray}
where $\mathcal{T}$ and $\mathcal{S}$ are the temperature and the entropy of a thermal state in the CFT, respectively, as defined in (\ref{eqn:temp}) and (\ref{eqn:entropy}), and $t$ is the ensemble temperature. The integrand under consideration can be interpreted as the deviation between all possible states within the canonical ensemble and the dual thermal states. The extreme points of the thermal potential exhibit the equilibrium states corresponding to the thermal states
\begin{eqnarray}
    \frac{dU}{d\mathcal{S}}=0\rightarrow t=\mathcal{T}.
\end{eqnarray}

\begin{figure}[ht]
  \begin{subfigure}
  \centering
    \includegraphics[width=0.4\linewidth]{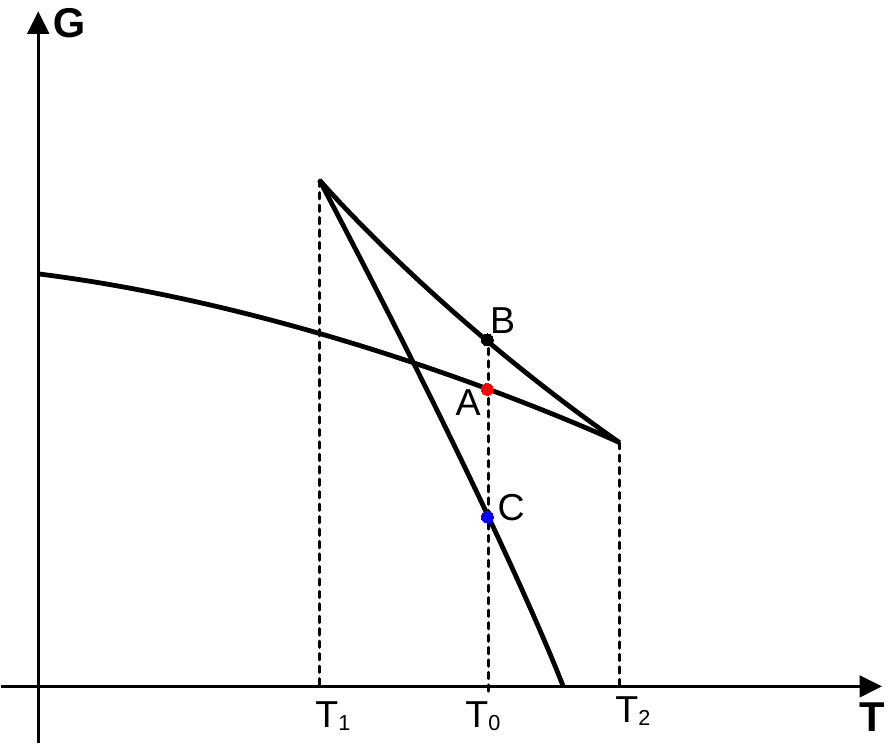}
  \end{subfigure}%
  \begin{subfigure}
  \centering
    \includegraphics[width=0.4\linewidth]{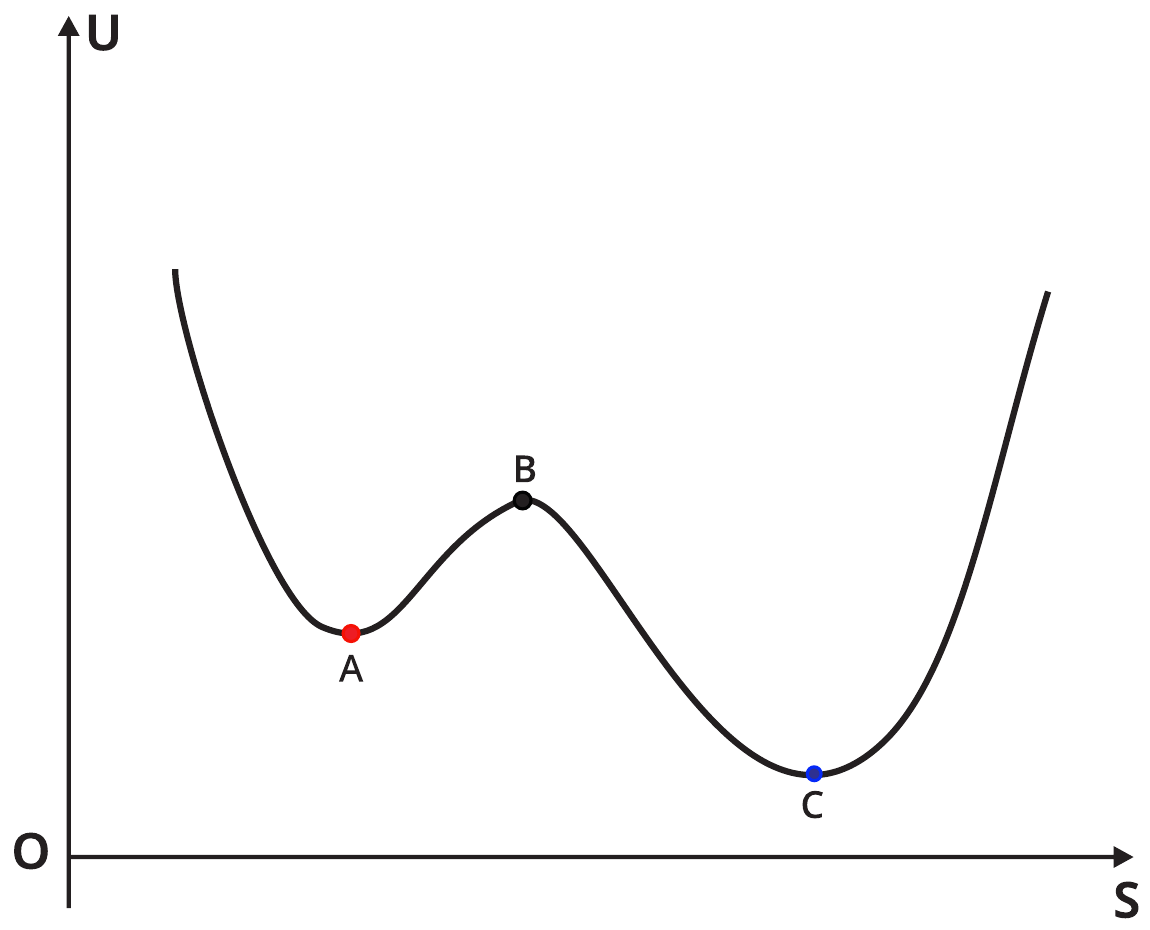}
  \end{subfigure}
  \caption{The schematic diagram of the thermal potential in a Van der Waals-like phase transition is illustrated. The left panel displays the $G-\mathcal{T}$ curve, where points $A,B$, and $C$ represent three states with the same temperature $\mathcal{T}_0$, satisfying $\mathcal{T}_1<\mathcal{T}_0<\mathcal{T}_2$. The right panel presents the $U-\mathcal{S}$ curve at temperature $\mathcal{T}_0$, with these states corresponding to the extreme points.}
  \label{fig:potential}
\end{figure}

The behavior of the thermal potential $U$ in a Van der Waals-like phase transition can be depicted in Fig. \ref{fig:potential}. Points $A,B$ and $C$ represent three thermal states at the same temperature $\mathcal{T}_0$. Points $A$ and $C$ correspond to local minima on the $U-\mathcal{S}$ curve, indicating stable thermal states. Point $B$ represents a local maximum on this curve, signifying an unstable thermal state. A thermal state at point $A$ can overcome the potential barrier at point $B$ and transition to point $C$. Another state can also revert from $A$ to $C$. The thermal states behave as Brownian particles in the potential $U$. We observe that the transition rates depend on the barrier heights $E_{AB}=U(x_B)-U(x_A)$ and $E_{CB}=U(x_B)-U(x_C)$. Assuming that the barrier heights are significantly larger than the temperature, the transition rates from $A$ to $C$ and from $C$ to $A$ can be expressed using the Kramer's escape rate formula, respectively,
\begin{eqnarray}
    r_{ac}=\frac{\sqrt{|U''(x_A)U''(x_B)|}}{2\pi}\exp{-\frac{U(x_B)-U(x_A)}{D}},\nonumber\\
    r_{ca}=\frac{\sqrt{|U''(x_C)U''(x_B)|}}{2\pi}\exp{-\frac{U(x_B)-U(x_C)}{D}},
    \label{eqn:rates}
\end{eqnarray}
where $D$ is the constant diffusion coefficient, which is proportional to the ensemble temperature \cite{Zwan2001}. We select the value of $D$ to ensure that the barrier heights are significantly greater than the temperature. However, if $D$ is too small, the exponential factors will become very large, causing the rates to approach zero. Thus, an appropriate value for $D$ is chosen to be 0.01.

\begin{figure}[ht]
  \begin{subfigure}
  \centering
    \includegraphics[width=0.3\linewidth]{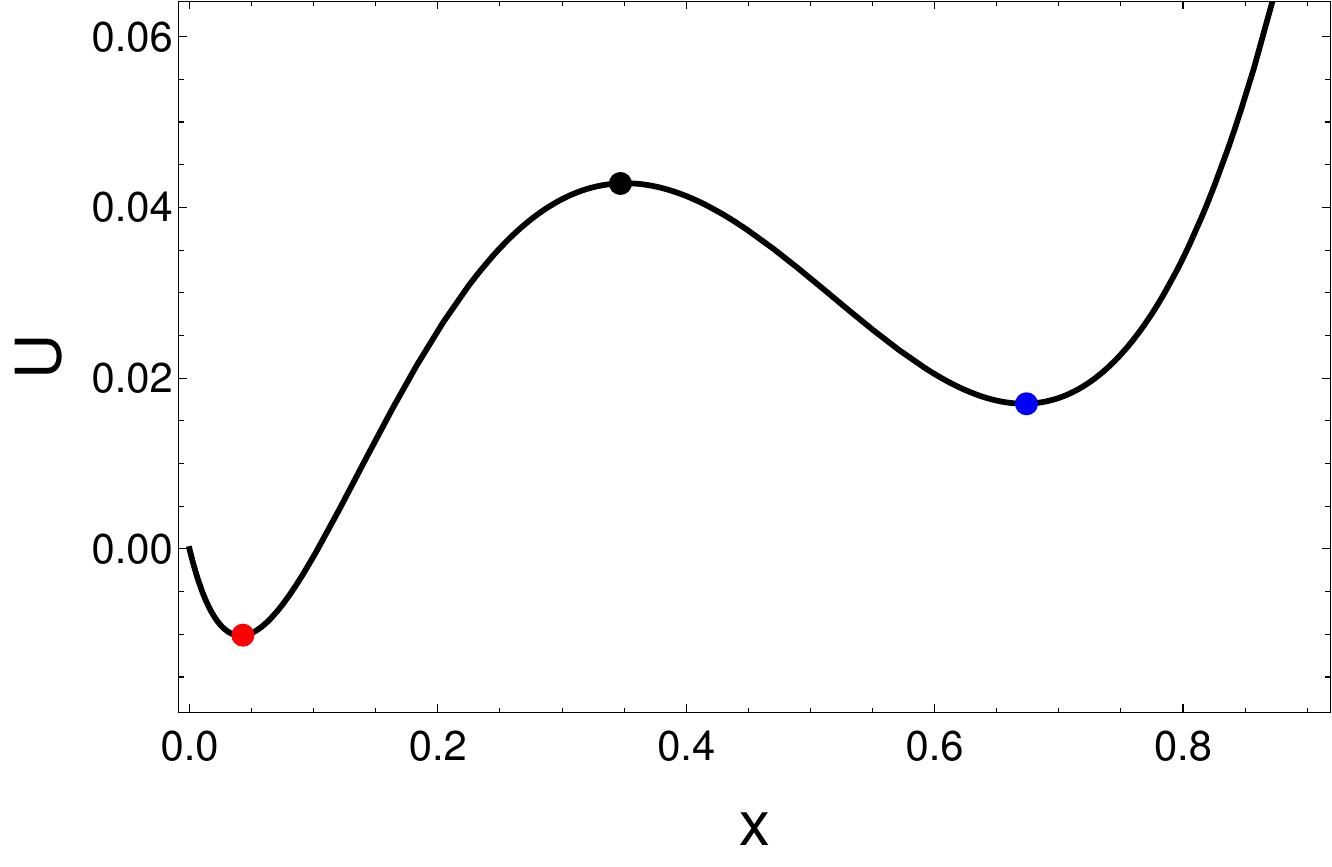}
  \end{subfigure}%
  \begin{subfigure}
  \centering
    \includegraphics[width=0.3\linewidth]{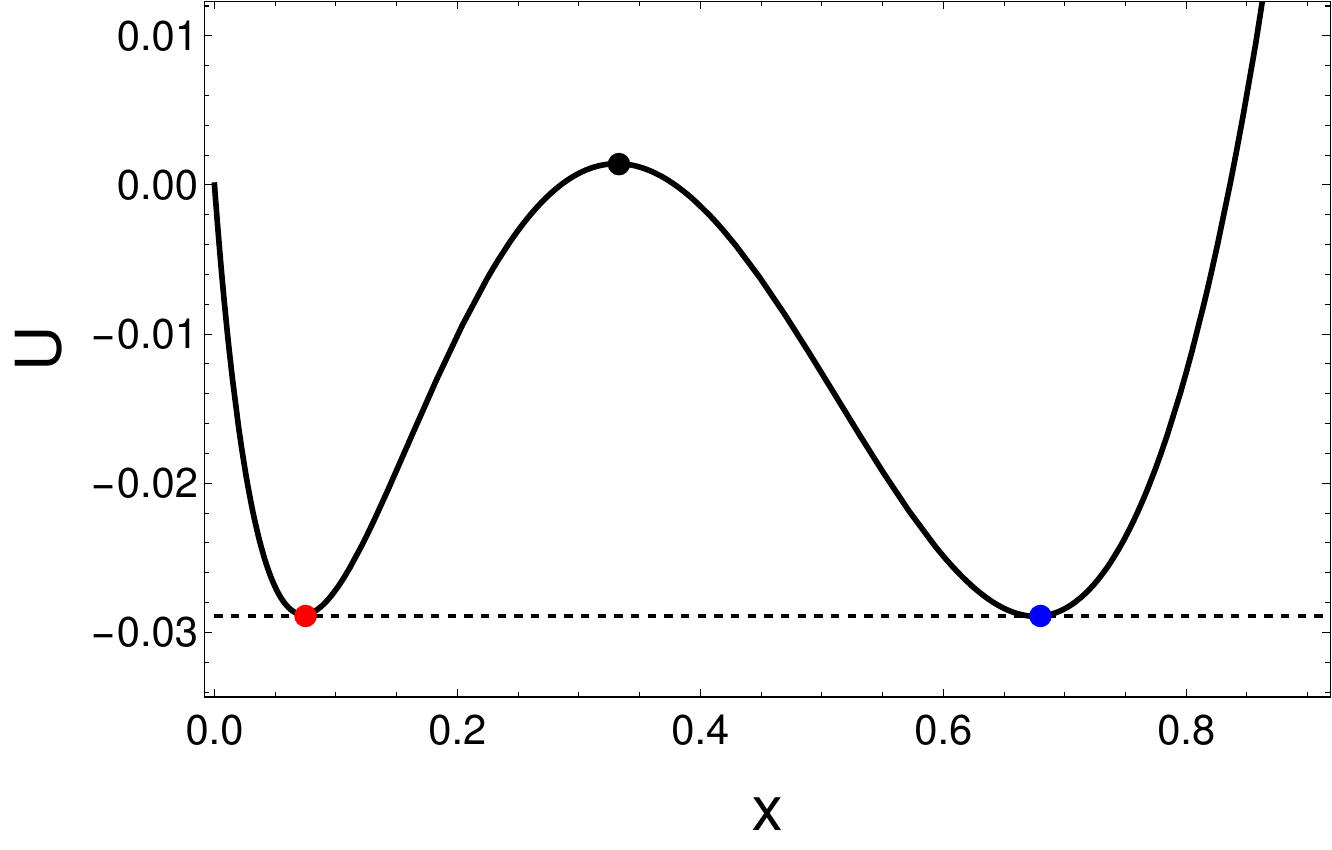}
  \end{subfigure}
  \  \begin{subfigure}
  \centering
    \includegraphics[width=0.3\linewidth]{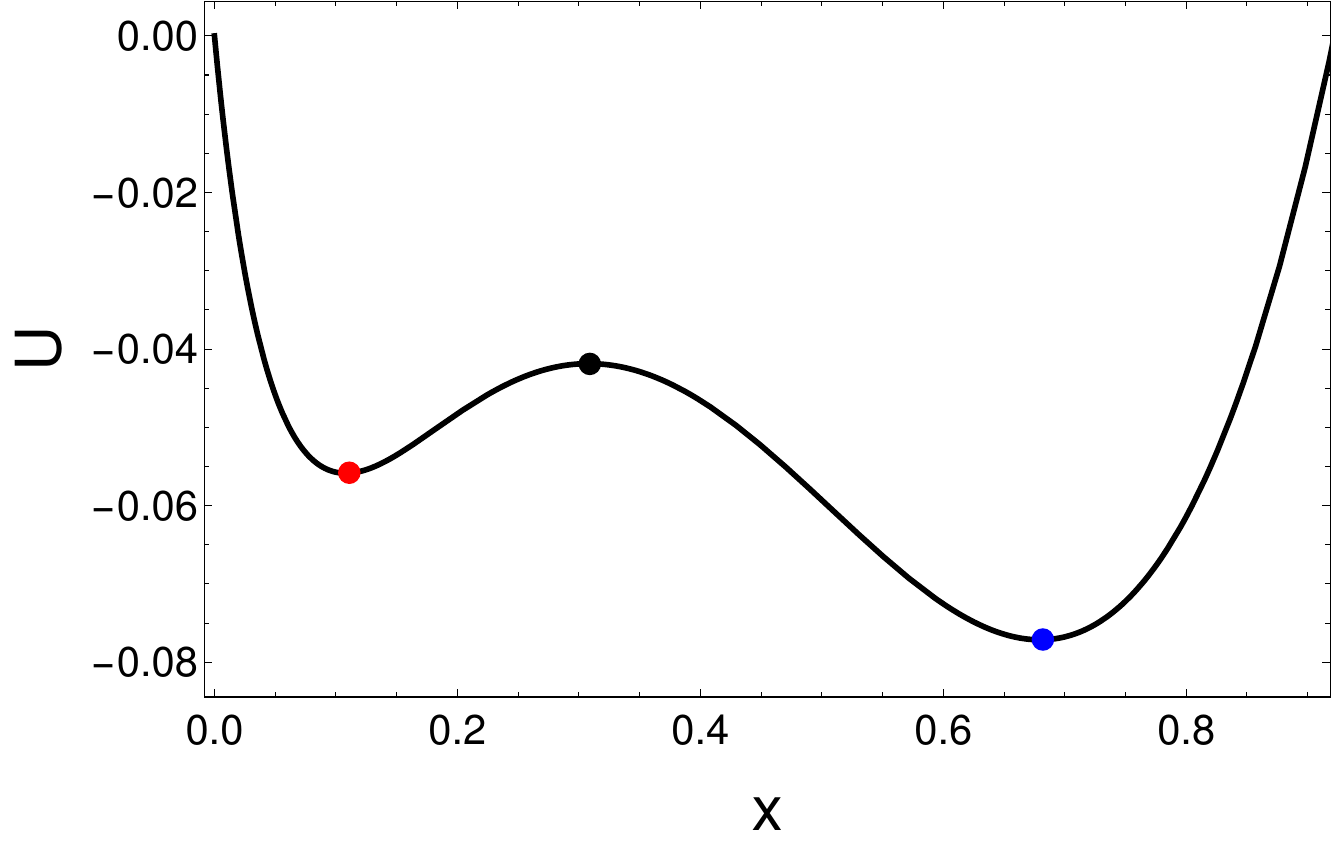}
  \end{subfigure}
  \caption{The thermal potential at $t=\mathcal{T}_0=0.27$. The left, middle, and right panels show the behavior of the thermal potential for $y<y_0$, $y=y_0$, and $y>y_0$, respectively.}
  \label{fig:potential1}
\end{figure}

From equations (\ref{eqn:temp}), (\ref{eqn:entropy}), and (\ref{eqn:therpot}), by fixing $R=1,C=1$, and $\mathcal{Q}=1$, we can integrate over the variable $x$ to derive the thermal potential formula as
\begin{eqnarray}
    U=-4\pi t x^2+x\left(-y^2+2(1+x^2+2xy^3)-8x^2y^4\log\left[1+\frac{1}{2xy}\right]\right).
\end{eqnarray}
In the previous section, we observed that the free energy $G$ exhibits a swallowtail shape when the dimensionless parameter $y$ satisfies the condition $y_{c_1}<y<y_{c_2}$. Additionally, for each consistent value $\mathcal{T}_0$, there exist two values $y_{min}$ and $y_{max}$ within the range $[y_{c_1},y_{c_2}]$ such that a Van der Waals-like phase transition occurs for $y_{min}<y<y_{max}$. For numerical calculations, given $\mathcal{T}_0=0.27$, we receive $y_{min}=y_{c_1}$ and $y_{max}\approx 2.199$. The behavior of the thermal potential at $t=\mathcal{T}_0$ is depicted in Fig. \ref{fig:potential1}. This result demonstrates that the relative height between the two local minima changes as $y$ varies. The left and right local minima represent the low and high entropy states, respectively. The local maximum corresponds to the unstable state, which rapidly transitions to the remaining states. When $y=y_0\approx 1.761$, the two minima in the thermal potential are equal. However, for $y<y_0$, the left minimum is lower than the right minimum, indicating that the low entropy state is more stable than the high entropy state. Conversely, for $y>y_0$, the right minimum is lower than the left minimum, indicating that the high entropy state is more stable than the low entropy state.

\begin{figure}[ht]
  \centering
    \includegraphics[width=0.5\linewidth]{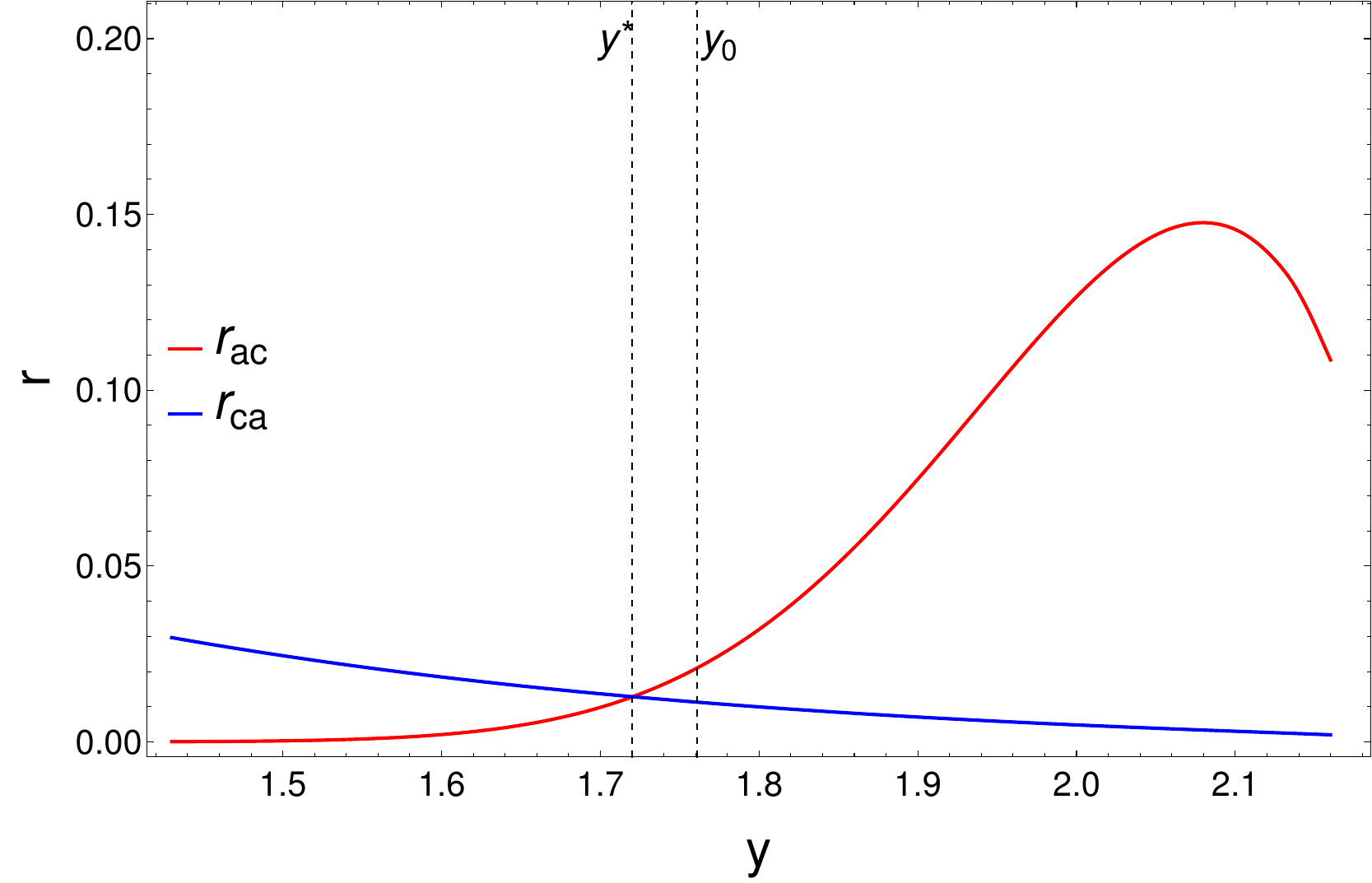}
  \caption{The phase transition rates at $t=0.27$. The solid red curve ($r_{ac}$) represents the rate of transition from low entropy states to high entropy states, while the solid blue curve ($r_{ca}$) depicts the rate of transition from high entropy states to low entropy states.}
  \label{fig:rate}
\end{figure}

We numerically calculate the two rates $r_{ac}$ and $r_{ca}$ in Equations (\ref{eqn:rates}) as $y$ varies within the range $y_{min}<y<y_{max}$. The results are presented in Fig. \ref{fig:rate}. For small values of $y$, the rate $r_{ac}$ is very low because the barrier height $E_{AB}$ is large, as shown in the left panel of Fig. \ref{fig:potential1}. Conversely, the rate $r_{ca}$ is higher than $r_{ac}$ because the barrier height $E_{CB}$ is smaller than $E_{AB}$. This result indicates that the transition process from the high entropy state to the low entropy state dominates the reverse process for small $y$. At $y=y*\approx 1.71$, the two rates are equal, despite $E_{AB}$ still being larger than $E_{CB}$. The barrier heights only become equal at $y=y_0>y*$, as depicted in the middle panel of Fig. \ref{fig:potential1}. This can be explained by the second-order derivatives in Equations (\ref{eqn:rates}). In other words, the transition rates also depend on the curvature of the potential at points $A$, $B$, and $C$. For $y>y*$, the rate $r_{ac}$ surpasses $r_{ca}$, rapidly increases to a maximum, and then decreases. Meanwhile, $r_{ca}$ gradually decreases to zero as $y$ increases. We can conclude that the transition process from the low entropy state to the high entropy state dominates the reverse process for large values of $y$.

\section{\label{sec:conclu} Conclusions}
Using the holographic dictionary from the proposal that treats Newton's constant as a thermodynamic variable, we find a thermodynamic topological equivalence between the AdS black holes in the bulk and the thermal states in the dual CFT. Additionally, we employ Duan's topological current theory to derive the topological winding numbers from the behavior of the generalized free energy. Assuming smoothness for the generalized free energy, we confirm the conjecture proposed in \cite{Weilima2022} that the global winding number only takes three values: -1, 0, and 1.

Our findings indicate that the characteristic parameter $\alpha$ of the regularized Maxwell theory significantly influences the topological features of RegMax AdS black holes. For $\alpha \leq \alpha_c = \frac{1}{\sqrt{2}}$, the local winding numbers are 1 and -1, corresponding to stable and unstable thermal states, respectively, resulting in a global winding number of zero. For $\alpha \geq \alpha_l$, there is only a stable state with a winding number of 1. In the remaining case, $\alpha_c < \alpha < \alpha_l$, the local winding numbers for the metastable, unstable, and stable states are 1, -1, and 1, respectively, yielding a global winding number of 1.

Employing the holographic dictionary, we investigate the phase transition in a canonical ensemble of the dual thermal states with fixed values of $(\mathcal{Q}, \mathcal{V}, C)$, considering its dependence on the characteristic parameter of the regularized Maxwell theory. The results indicate the existence of a Van der Waals-like phase transition when $y$ lies between two critical constant values. Treating the phase transition as a stochastic process, we quantitatively calculate the Kramer escape rates between the stable thermal states. Our observations reveal that these rates are closely related to the behavior of the thermal potential. This finding provides a clearer understanding of the dominant processes in the phase transition between the low and high entropy thermal states in the dual CFT.

\end{document}